\documentclass[pra,10pt,twocolumn,superscriptaddress,floatfix,showpacs]{revtex4-1}
\usepackage{subfig}
\usepackage{graphicx} 
\usepackage{epstopdf}
\usepackage[utf8]{inputenc}
\usepackage[T1]{fontenc}     
\usepackage[british]{babel}  
\usepackage[sc,osf]{mathpazo}
\usepackage[scaled=0.86]{berasans}  
\usepackage[colorlinks=true, allcolors=blue]{hyperref}  
\usepackage[babel]{microtype}  
\usepackage{amsmath,amssymb,amsthm,bm,amsfonts,mathrsfs,bbm} 

\usepackage{xspace}  
\usepackage{pgfplots}
\usepackage{xcolor,colortbl}
\def\ba{\begin{equation}}
\def\ea{\end{equation}}
\def\bea{\begin{eqnarray}}
\def\eea{\end{eqnarray}}
\def\ben{\begin{equation*}}
\def\een{\end{equation*}}
\def\bean{\begin{eqnarray*}}
\def\eean{\end{eqnarray*}}
\def\bma{\begin{mathletters}}
\def\ema{\end{mathletters}}
\def\bi{\begin{itemize}}
\def\ei{\end{itemize}}

\newcommand{\be}{\begin{equation}}
\newcommand{\ee}{\end{equation}}

\newcommand{\kommentar}[1]{}

\newcommand{\forget}[1]{}

\begin{document}

\title{Generation of Nonlocality}

\author{Kaushiki Mukherjee}
\email{kaushiki_mukherjee@rediffmail.com}
\affiliation{Department of Mathematics, Government Girls’ General Degree College, Ekbalpore, Kolkata-700023, India.}

\author{Biswajit Paul}
\email{biswajitpaul4@gmail.com}
\affiliation{Department of Mathematics, South Malda College, Malda, West Bengal, India}

\author{Debasis  Sarkar}
\email{dsappmath@caluniv.ac.in, dsarkar1x@gmail.com}
\affiliation{Department of Applied Mathematics, University of Calcutta, 92, A.P.C. Road, Kolkata-700009, India.}

\author{Amit Mukherjee}
\email{amitisiphys@gmail.com}
\affiliation{Physics and Applied Mathematics Unit, Indian Statistical Institute, 203,B. T. Road, Kolkata 700108 , India.}

\author{Some Sankar Bhattacharya}
\email{somesankar@gmail.com}
\affiliation{Physics and Applied Mathematics Unit, Indian Statistical Institute, 203,B. T. Road, Kolkata 700108 , India.}

\author{Arup Roy}
\email{arup145.roy@gmail.com}
\affiliation{Physics and Applied Mathematics Unit, Indian Statistical Institute, 203,B. T. Road, Kolkata 700108 , India.}
\author{Nirman Ganguly}
\email{nirmanganguly@gmail.com}
\thanks{On leave from Department of Mathematics, Heritage Institute of Technology, Kolkata-107,India}
\affiliation{Physics and Applied Mathematics Unit, Indian Statistical Institute, 203,B. T. Road, Kolkata 700108 , India.}


\begin{abstract}
Environmental influences are typical in any practical situation which in turn can have fatal effects on quantum resources. Bell nonlocality is such an important resource. Some environmental interactions can lead to nonlocality being lost. In such situations, it is vital to find possible prescriptions to retrieve nonlocality. The present work lays down one such prescription. Precisely, we have studied some well-known classes of states under the ambit of the Bell-CHSH inequality in two qubits, where we start from a Bell-CHSH local state and can transform it into a nonlocal state through our protocol. The efficacy of the protocol is further established from the fact that it can retrieve nonlocality from states admitting a LHV(local hidden variable) model. The strength of the prescription is validated by the fact that it can generate nonlocality from states when even unitary action on the composite system fails.
\end{abstract}

\maketitle
	
\section{Introduction}
Quantum nonlocality is generally interpreted as the ability of quantum theory to explain correlations arising due to local measurements on space-like separated systems beyond the scope of local realism. Since its origin from Bell's theorem \cite{Bell64,CHSH69}, it has been the cynosure of significant research activity. Entanglement \cite{Horodecki'2009} marks another significant departure of quantum mechanics from its classical counterpart. Although inequivalent in concept, nonlocality and entanglement has proved themselves to be pertinent resources in various tasks like teleportation \cite{Bennett93}, randomness certification \cite{random}, cryptography \cite{Bennett84}. While entanglement refers to inseparability of subsystems, Bell-nonlocality is interpreted as the strongest form of quantum nonlocality. Bell-nonlocality is detected through the violation of a suitable Bell inequality, one such important inequality being the CHSH inequality\cite{CHSH69}, which itself is necessary and sufficient under two measurement settings. A state which satisfies the Bell-CHSH inequality cannot be considered \textit{local}, as there can be other Bell inequalities that it violates. However, violation of the Bell-CHSH inequality is a signature of nonlocality. A state is termed as \textit{local}, only if it admits a local hidden variable model(LHV) \cite{fine82}. The correlations exhibited by a state admitting a LHV can be simulated by separable states. The inequivalence of entanglement and nonlocality is exemplified by the existence of entangled states admitting a LHV \cite{werner89}. \\
\indent The underlying foundational and pragmatic significance of nonlocality has motivated research towards revealing nonlocality\cite{Brunner2014}, local filtering operation being one such procedure \cite{popescu95,gisin96}. The procedures to reveal hidden nonlocality can be broadly classified into two categories (i) performing single local measurement \cite{gisin96} and (ii) subjecting the state to suitable sequence of local measurements \cite{popescu95}. There has been extensive research in both the directions, the main motivation being extraction of nonlocal behaviour from quantum states which failed to generate the same under the standard Bell scenario. Recently, the effect of global unitary operations on the nonlocality of a state has been probed by some of us \cite{ganguly16}, with the focus being on the Bell-CHSH inequality for two qubit systems. Precisely, a state which initially satisfies the Bell-CHSH inequality can violate it when subjected to a global unitary action. On the other hand, there are states which preserve their Bell-CHSH local character under global unitary effect, subsequently termed as \textit{absolutely Bell-CHSH local} states \cite{ganguly16,ganguly2}. For a given spectrum, the maximal Bell-CHSH violation is attained at the respective Bell-diagonal state\cite{wolf2002}. Therefore, if a Bell-diagonal state is Bell-CHSH local, it is absolutely Bell-CHSH local\cite{ganguly16}. Hence, a quantum state with spectrum $ a_1,a_2,a_3,a_4 $(in decreasing order) is termed as absolutely Bell-CHSH local iff \cite{ganguly2},
\begin{equation}
(2a_1+2a_2-1)^2+(2a_1+2a_3-1)^2 \le 1
\end{equation}
This feature bears resemblance to a similar problem in entanglement, where there are states which preserve separability under global unitary action on the composite system\cite{absep zyck,absep ver,johnston}. Such states are termed as absolutely separable.\\
\indent On a different note, a state may lose its nonlocality if it is subjected to environmental interaction, a situation one cannot dispense within practical scenarios\cite{ghosh13}. Recent studies on nonlocality breaking maps have been indicative of the probable drastic effects\cite{ghosh13}. In this context, one pertinent question is whether one can regenerate nonlocality. The question assumes significance if one can retrieve nonlocality even from absolutely Bell-CHSH local states.\\
\indent Given the practical and foundational significance of nonlocality, a quantum state capable of exhibiting the trait can be considered to be a resource.Now if nonlocality of any entangled state does not appear by using existing procedures of detection, the state may seem useless in spite of being entangled. Hence from a resource theoretic perspective, it becomes important to generate nonlocality from such useless entangled states. In recent times there has been many studies attempting to exhibit nonlocal features of entangled quantum states which are thought to be devoid of any capability to generate nonlocal correlations in standard Bell sense \cite{Bell64,CHSH69}.\\
\indent Our present work is motivated by the aforementioned questions. To state precisely, we find that an amplitude damping channel, a typical representative of environmental interaction, can have fatal effects on the nonlocality of a state. The state can be even turned into an absolutely Bell-CHSH local state. However, through appropriate sequential measurements one can transform the state to a state which is no longer absolutely Bell-CHSH local. One can then generate nonlocality from the transformed state through a suitable global unitary. The protocol relies on the SLOCC(Stochastic Local Operation and Classical Communication) used in the standard entanglement swapping procedure \cite{ekert1993}. We have chosen as input the X class of states \cite{eberlyqic}, which contains within it several significant subclasses of quantum states. Our protocol is further underscored by existence of absolutely Bell-CHSH local states admitting a LHV model which can be converted to a state which is no longer absolutely Bell-CHSH local.\\
\indent We have organized the remaining work as follows: In the following section we observe the influence of an amplitude damping channel on the Werner class of states. In sec.\ref{cs1} we introduce the protocol and prescribe measures to retrieve nonlocality. We also probe the generation of nonlocality from well-known class of states in sec.\ref{cs2}. We finally conclude in sec.\ref{concl}.
\section{Amplitude Damping}\label{ampdamp}
In this section we give an intuitive insight of our protocol and its possible applications before presenting the main prescription in the following section.
Consider the Werner class of states \cite{werner89}:
\begin{equation}\label{werner}
\Lambda=\alpha |\psi^-\rangle\langle\psi^-|+(1-\alpha)\frac{\mathbf{1}}{4},\,\textmd{where}\,\alpha\in[0,1].
\end{equation}
After being passed through an amplitude damping channel(characterized by noise parameter $\gamma$,say)\cite{Nielsen}, the transformed state is given by:
\begin{widetext}
\begin{equation}\label{ampd}
  \Lambda_{\textmd{AMP}}=\left(\begin{array}{cccc}
\frac{(1-\alpha)(1+\gamma^2)+2(1+\alpha)\gamma}{4}&0&0&0\\
0&\frac{(1-\gamma)(1+\gamma+\alpha(1-\gamma))}{4}&-\frac{\alpha(1-\gamma)}{2}\\
0&-\frac{\alpha(1-\gamma)}{2}&\frac{(1-\gamma)(1+\gamma+\alpha(1-\gamma))}{4}&0\\
0&0&0&\frac{(1-\alpha)(1-\gamma)^2}{4}\\
\end{array} \right)
\end{equation}
\end{widetext}

\begin{figure}[t]
\centering
\includegraphics[width=1.5in]{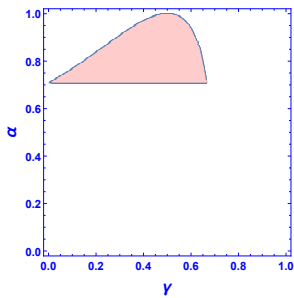}
\caption{\emph{Shaded region gives a restricted area in parameter space $(\alpha,\gamma)$ where $\alpha$ and $\gamma$ characterizes the Werner state and amplitude damping channel respectively. A Werner state corresponding to any point say $P$ in this restricted area loses its nonlocal character when passed through an amplitude damping channel parametrized by noise parameter $\gamma$ corresponding to the point $P.$ The extent of damping is such that its nonlocal behavior cannot be extracted even by applying optimal global unitary operations.}}
\label{amplitude1}
\end{figure}

The nonlocal states from Werner class after passing through amplitude damping channel become absolutely Bell-CHSH local, i.e., nonlocal character of the damped states cannot be retrieved even under global unitary operations(see Fig.(\ref{amplitude1})).

\begin{figure}[t]
\centering
\includegraphics[width=1.5in]{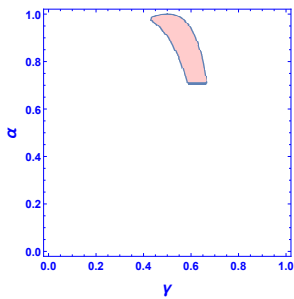}
\caption{\emph{Shaded region gives restrictions for which, via our swapping protocol, nonlocality(under suitable global unitary operations) can be retrieved(if Bob obtains output $|\psi^{+}\rangle$ or $|\psi^{-}\rangle$) from absolutely Bell-CHSH local Werner states resulting from amplitude damping channel parametrized by $\gamma$ lying in the shaded region.}}
\label{amplitude2}
\end{figure}

However when such states are used in our protocol then nonlocal character of some of them can be retrieved probabilistically in the measurement phase of the protocol when suitable global unitary operations are applied on the states resulting due to Bob's joint measurement(see Fig.(\ref{amplitude2})).
The parameter range for the transformation and subsequent retrieval is given below:\\
\begin{center}
\begin{table}[htp]
  	\begin{center}
  		\begin{tabular}{|c|c|c|c|c|c|}
  			\hline
  			Noise &Absolutely&Non- absolutely &Range of\\
  Parameter&  Bell-CHSH& local swapped&retrieving \\
    &  local&  state& nonlocality\\
  			\hline
  			$\gamma=0.5$&$\alpha\leq 1$&$\alpha>0.908433 $&$(0.908433,1]$\\
  			\hline
  		\end{tabular}\\
  		\caption{}
  	\end{center}
  	\label{table1}
  \end{table}
  \end{center}
   \vspace{0.2 cm}
\section{Sequential Measurement Protocol}\label{cs1}
Before moving into the description of the main protocol, we discuss briefly two key ingredients.\\
\subsection{Absolutely Bell-CHSH local states}
One very pertinent question in the context of the Bell-CHSH inequality is what can be the maximal Bell-CHSH violation of a state under global unitary operations. It was shown in \cite{wolf2002}, that if one fixes the spectrum of a state then the maximal violation is attained at the respective Bell-diagonal state. Therefore , if the respective Bell-diagonal state do not show any violation then the states unitarily similar to that Bell-diagonal state are absolutely Bell-CHSH local \cite{ganguly16}. This puts a restriction on the spectrum of the state. Precisely, a quantum state with spectrum $ a_1,a_2,a_3,a_4 $(in decreasing order) is termed as absolutely Bell-CHSH local iff \cite{ganguly2},
\begin{equation}
\label{abloc}
(2a_1+2a_2-1)^2+(2a_1+2a_3-1)^2 \le 1
\end{equation}
The above relation can also be obtained if one considers the Bell-CHSH violation of the Bell diagonal state for a given spectrum \cite{wolf2002}.
\subsection{X states}
This class of bipartite states\cite{eberlyqic} are of the form:

\begin{eqnarray}\label{st9}
\chi=a|00\rangle \langle 00|+b|01\rangle\langle 01|+c|10\rangle\langle10|
+d|11\rangle\langle11|\nonumber\\
+p(|00\rangle\langle11|+|11\rangle\langle00|)+q(|01\rangle\langle10|+|10\rangle\langle01|)
\end{eqnarray}

Here$ p$ and $q$ real, $a+b+c+d=1$. Non-negativity demands $p^2\leq a d$ and $q^ 2\leq bc$. The corresponding density matrix is given by:
\begin{equation}\label{x}
\rho_{\chi}=
   \left(\begin{array}{cccc}
      a &0 & 0 & p\\
      0 & b & q & 0\\
      0 & q & c & 0\\
      p &0 &0 & d \\
       \end{array} \right)
\end{equation}
The X states are well known for their versatile utility in experimental scenarios\cite{Rau,Bose}. This class includes many well known class of states such as Bell diagonal states, Werner state\cite{werner89}. X states were also studied in condensed matter systems and in various other fields of quantum mechanics.
\subsection{The Protocol}
\begin{center}
\begin{figure}[htb]
	\centering
	\includegraphics[width=3.5in]{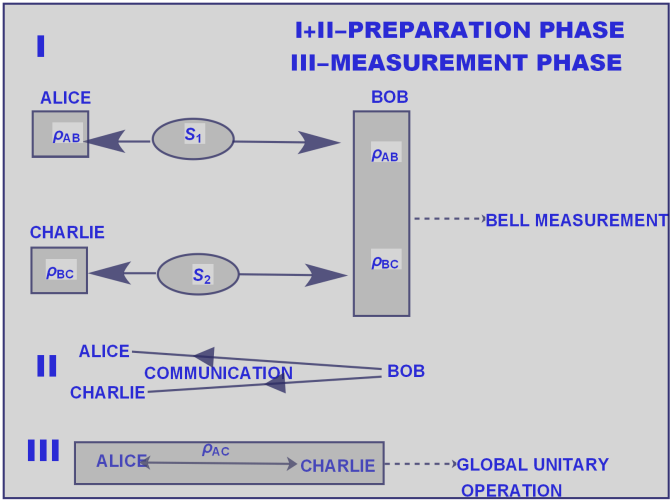}
\caption{\emph{Schematic diagram of the protocol}}
\label{pro}
\end{figure}
\end{center}

The protocol that we put to use mainly relies on an entanglement swapping network which entangles two never interacting pair of particles through suitable measurements. The protocol consists of two phases: \textit{Preparatory Phase} followed by \textit{Measurement Phase}. Details of the protocol(see Fig.(\ref{pro})) are discussed below:\\
\indent Let there be three parties Alice, Bob and Charlie and two sources $S_1$ and $S_2$. Each of the two sources generates an entangled state. Let $S_1$ generate $\rho_{AB}$ which is shared between Alice and Bob. Let $\rho_{BC}$ be generated by $S_2$ and shared between Bob and Charlie. So Bob receives two particles, one from each source. After receiving two particles, he performs full Bell-basis measurement on the joint state of his two particles. He then broadcasts the output of his joint measurements to Alice and Charlie. This constitutes the first phase, i.e., \textit{Preparatory Phase}. \\
\indent After receiving the output from Bob, Alice and Charlie now know the state that they share. They perform suitable local projective measurements on their subsystems. The correlations generated in this phase are tested to see whether they are absolutely Bell-CHSH local(Eq.(\ref{abloc})). This is the \textit{Measurement Phase} of the protocol.\\
The aforementioned protocol is now put to use in the context of generating nonlocality. For that let each of two sources $S_1$ and $S_2$ generates an absolutely local state $\rho_{AB}$ and $\rho_{BC}$ respectively. After completion of the sequence of measurements in the protocol, if the output state is not absolutely Bell-CHSH local(which is revealed through the violation of Eq.(\ref{abloc})), then a suitable unitary is applied on it to change it into a non-local state. \\
\indent Our protocol, being an SLOCC, nonlocality is generated with some non zero probability, i.e., depending on Bob obtaining some specific output, the conditional bipartite state shared between Alice and Charlie is not absolutely Bell-CHSH local. In this context, efficacy of the protocol can be considered to be further improved if via this protocol one can generate states which are not absolutely Bell-CHSH local irrespective of Bob's output. Interestingly, for some states, we have the answer in affirmative.

\section{Observations}\label{cs2}
First we consider both $\rho_{AB}$ and $\rho_{AC}$ to be absolutely Bell-CHSH local.
\subsection{Nonlocality from absolutely Bell-CHSH local states}
The condition for X states to be absolutely Bell-CHSH local translates to,
\begin{equation}
\label{al1}
    L_{abs}\leq1
\end{equation}
where $L_{abs}$ is given by:
\begin{eqnarray}
\label{al2}
L_{abs}=(\Theta_1+\Theta_2)^2+\Theta_3^2\,&\,\textmd{if}\,&\,\Theta_3^2\geq (\Theta_1-\Theta_2)^2
,\nonumber \\ \Theta_1^2+\Theta_2^2\,&\,\textmd{if}\,&\,\Theta_3^2\leq (\Theta_1-\Theta_2)^2\nonumber \\
\end{eqnarray}
with $\Theta_1=\sqrt{(a-d)^2+4p^2}$, $\Theta_2=\sqrt{(b-c)^2+4q^2}$ and $\Theta_3=a+d-b-c.$\\

Let two different copies of $X$ states $\chi_1$ and $\chi_2$ be generated by $S_1$ and $S_2$ respectively. Let $\chi_{nl}$ denote the $X$ state shared between Alice and Charlie based on Bob's output at the end of the preparation phase of our protocol. Below we consider some subclasses of $X$ state for which generation of nonlocality is observed. \\
\subsubsection{States diagonal in the Computational Basis}
\begin{center}
\begin{figure}[t]
\onecolumngrid
\begin{minipage}{.3\textwidth}
\begin{tabular}{|c|c|}
\hline
\subfloat[$\varphi_f^{00}$ or $\varphi_f^{01}$ ]{\includegraphics[trim = 0mm 0mm 0mm 0mm,clip,scale=0.3]{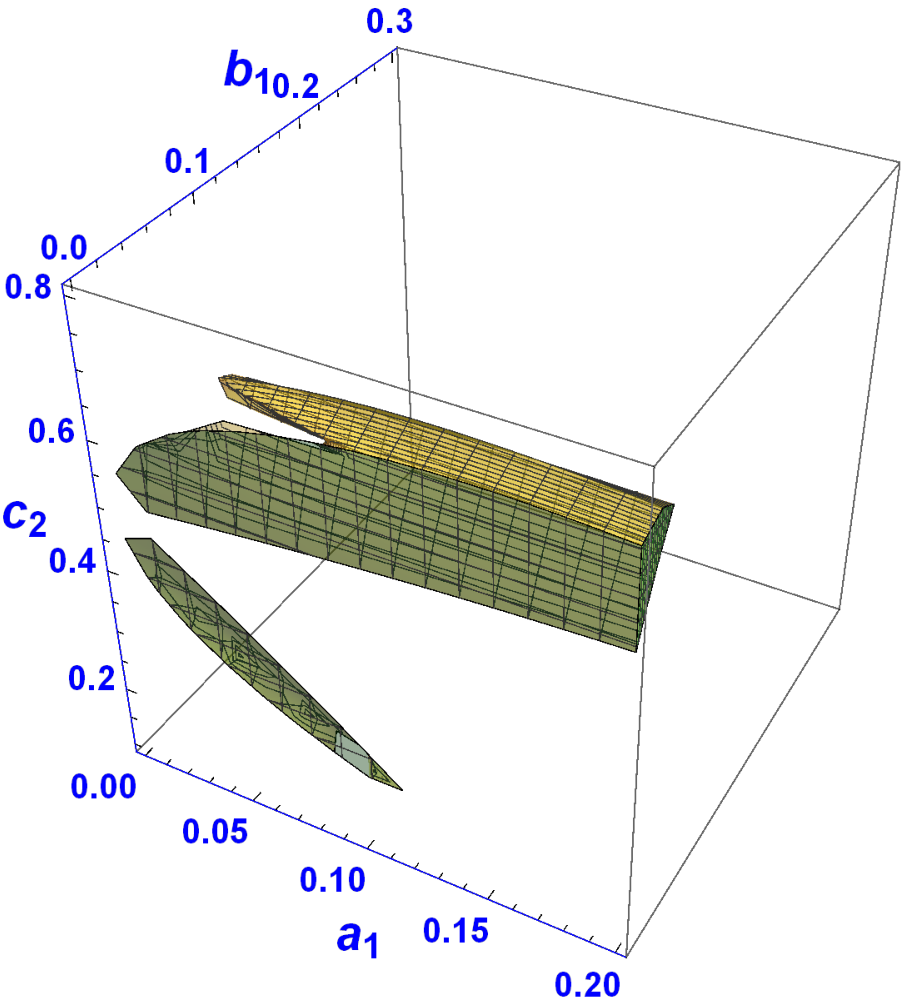}} &
\subfloat[$\varphi_f^{10}$ or $\varphi_f^{11}$]{\includegraphics[trim = 0mm 0mm 0mm 0mm,clip,scale=0.3]{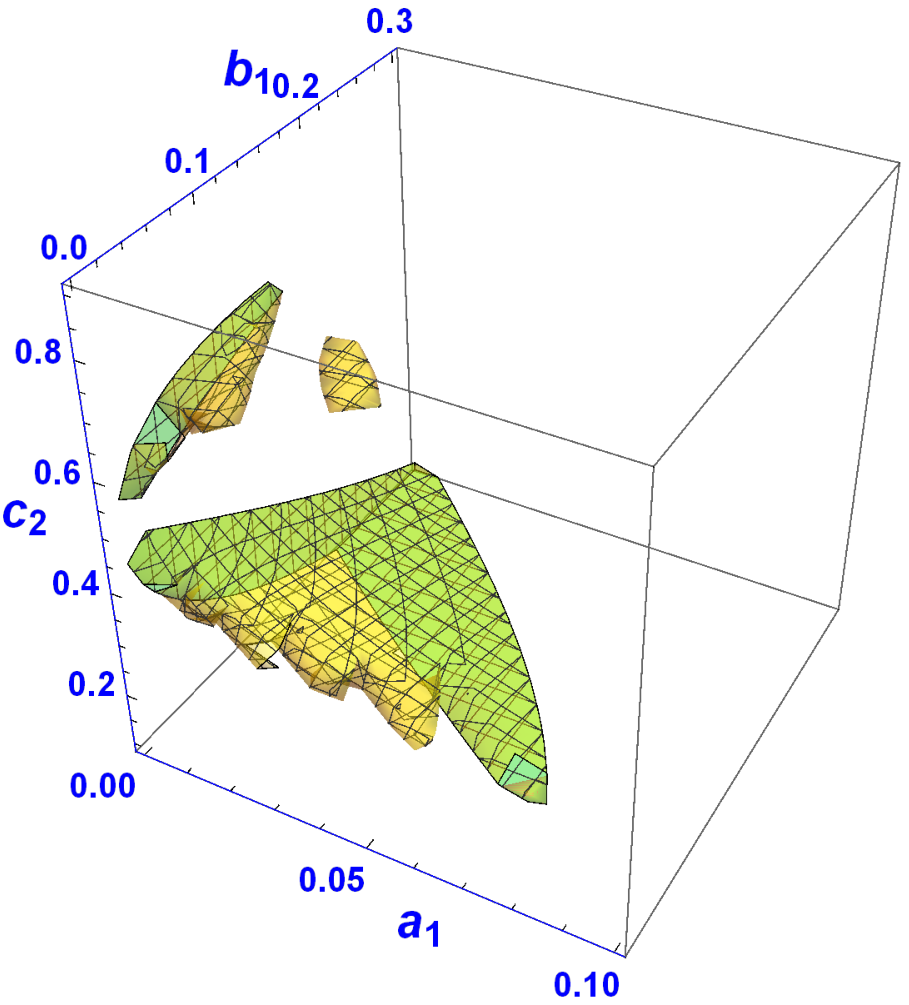}}\\
\hline
\end{tabular}
\caption{\emph{Both the subfigures give restrictions over state parameters of absolutely Bell-CHSH local $\varphi_1$ and $\varphi_2$ along with $a_1=a_2$, $b_1=b_2$ and $c_1=\frac{1}{2} $ for which non absolutely local states are generated at the end of the protocol.}}\label{abs3}
\end{minipage}
\begin{minipage}{.3\textwidth}
\begin{tabular}{|c|c|}
\hline
\subfloat[$\varphi_f^{00}$ or $\varphi_f^{01}$ ]{\includegraphics[trim = 0mm 0mm 0mm 0mm,clip,scale=0.3]{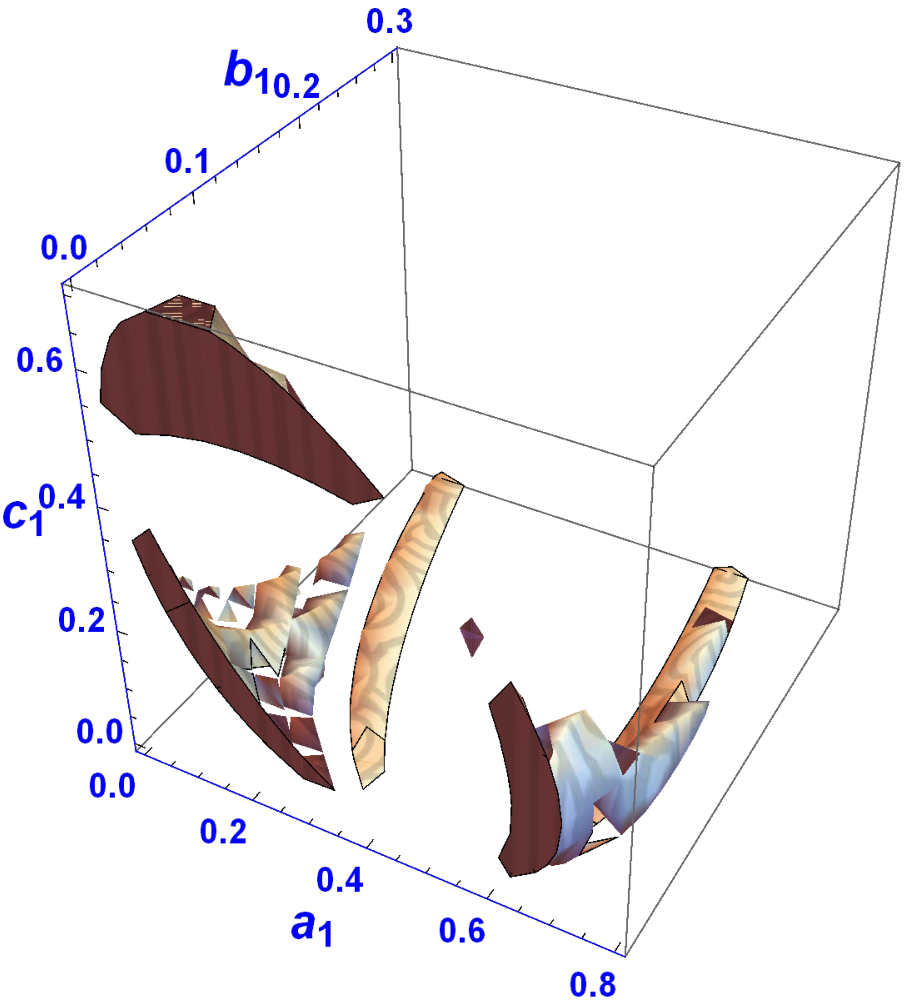}} &
\subfloat[$\varphi_f^{10}$ or $\varphi_f^{11}$]{\includegraphics[trim = 0mm 0mm 0mm 0mm,clip,scale=0.3]{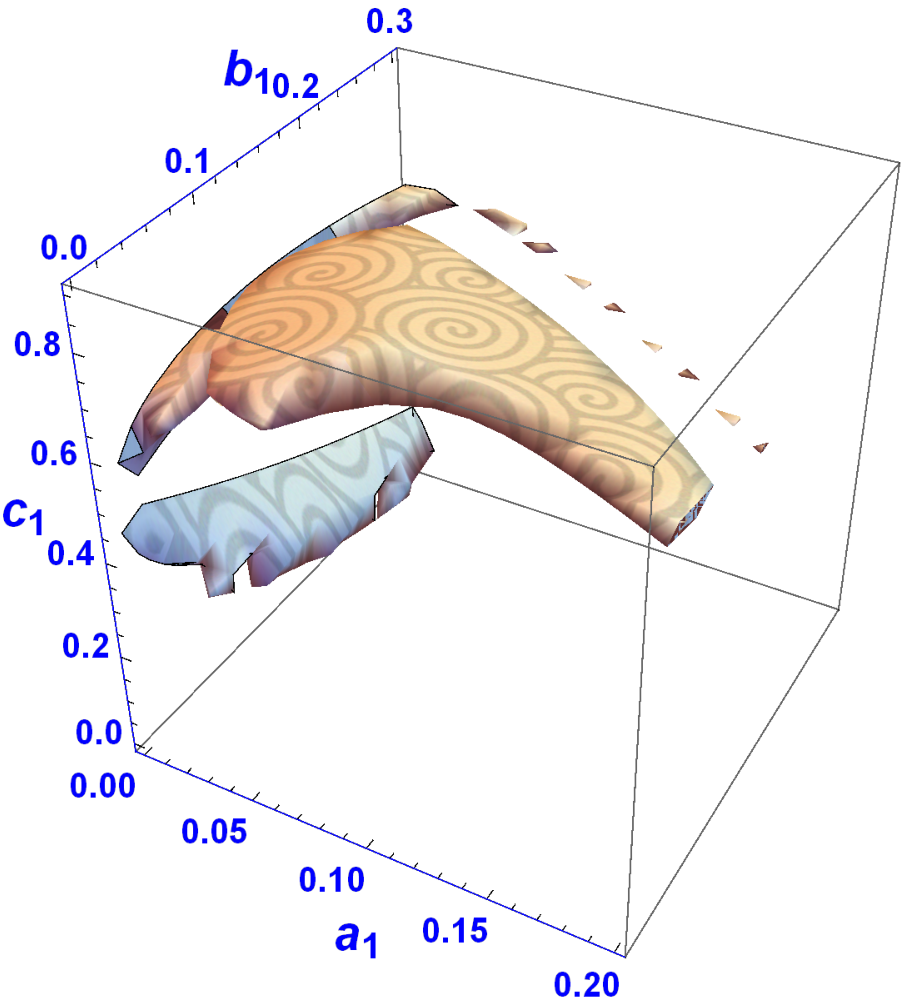}}\\
\hline
\end{tabular}
\caption{\emph{Shaded regions in both subfigures give restricted areas in subspace of the state parameter space ($a_1$, $b_1$, $c_1$) of $\varphi_1$(Eq.(\ref{cb1})) for which non absolute local states are obtained in the measurement phase of the protocol starting from identical copies of absolutely Bell-CHSH local states.}}\label{abs4}
\end{minipage}
\end{figure}
\end{center}

Consider $S_i$ generates a $X$ state which is diagonal in the computational basis($\{|00\rangle,|01\rangle,|10\rangle,|11\rangle\}$):
\begin{equation}\label{cb1}
    \varphi_i=a_i|00\rangle\langle 00|+b_i|01\rangle\langle01|+c_i|10\rangle\langle10|+d_i|11\rangle\langle11|
    \end{equation}
with $ i=1,2 $ , $a_i\in[0,1]$, $b_i\in[0,1]$, $c_i\in[0,1]$, $d_i\in[0,1]$ and $a_i+b_i+c_i+d_i=1.$ If both $\varphi_1$ and $\varphi_2$ used in the protocol be absolutely Bell-CHSH local(Eqs.(\ref{al1},\ref{al2}), let the swapped state resulting due to Bob obtaining outcome $b_{ij}$ be denoted as $\varphi_f^{ij}.$  For rest of our discussion, we denote the outputs of Bob, i.e., the Bell states $|\phi^{+}\rangle$, $|\phi^{-}\rangle$, $|\psi^{+}\rangle$ and $|\psi^{-}\rangle$ as $b_{00}$, $b_{01}$, $b_{10}$ and $b_{11}$ respectively. So for this class of states(Eq.(\ref{cb1})), there exist some absolutely local states from this family(Eq.(\ref{cb1})) which when used in our protocol, non absolutely Bell-CHSH local states are generated for any possible output of Bob. If Bob obtains output $b_{00}$ or $b_{01}$, $\phi_f^{00}$ or $\phi_f^{01}$ violate the absolute Bell-CHSH locality criteria(Eq.(\ref{al2})) for the same restrictions over the state parameters. Similarly if Bob obtains output $b_{10}$ or $b_{11}$, in both cases, the conditional state shared between Alice and Charlie, i.e., $\phi_f^{10}$ or $\phi_f^{11}$ become non absolutely Bell-CHSH local. For a particular instance, consider two copies $\varphi_1$ and $\varphi_2$ with $a_1$$=$$a_2$, $b_1$$=$$b_2$ and $c_1=\frac{1}{2}$. When used in our protocol, the states get transformed into non-absolutely Bell-CHSH local regime (see Fig.(\ref{abs3})).
\twocolumngrid
Now if the sources generate identical copies, then too one obtains states which are no longer absolutely Bell-CHSH local(see Fig.(\ref{abs4}))\\
Clearly there exists no common intersection region of the two subfigures of Fig.(\ref{abs4}) which indicates that here generation of non absolute locality depends on the output of Bob's Bell-basis measurement. To be specific there exist no such absolutely Bell-CHSH local state from this family(Eq.(\ref{cb1})) for which  generation is possible deterministically, i.e., irrespective of Bob's output.
\subsubsection{Gisin states}
Consider that $S_1$ and $S_2$ generates two different copies $\rho_1$ and $\rho_2$ respectively of the Gisin's state \cite{gisin96}:

\begin{eqnarray}\label{gis1}
    \rho_i&&=\lambda_i(\sin^2\alpha_i|01\rangle\langle01|+\cos^2\alpha_i|10\rangle\langle10|\nonumber\\&&+\sin\alpha_i\cos\alpha_i(|10\rangle\langle01|+|01\rangle\langle10|))\nonumber\\&&+\frac{1-\lambda_i}{2}(|00\rangle\langle00|+|11\rangle\langle11|)\nonumber\\&&\alpha_i\in[0,\frac{\pi}{4}],\,\lambda_i\in[0,1].
\end{eqnarray}

\begin{center}
\begin{figure}[htb]
	\centering
	\includegraphics[width=2.5in]{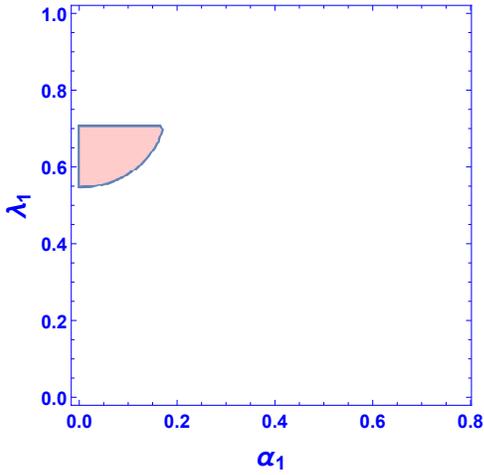}
\caption{\emph{The figure gives region in state parameter space ($\alpha_1$, $\lambda_1$) of Gisin states(Eq.(\ref{gis1})) for which non absolute Bell-CHSH locality is observed in our protocol when identical copies of Gisin states are used and Bob obtains output $|\phi^+\rangle(b_{00})$ or $|\phi^-\rangle(b_{01}).$ }}
\label{abs2}
\end{figure}
\end{center}

For any value of state parameter $\alpha_i$, $\rho_i$ is absolutely Bell-CHSH local($L_{abs}\leq1$) if $\lambda_i\in[0,\frac{1}{\sqrt{2}}]$. When absolutely Bell-CHSH local $\rho_i,\,i=1,2$(i.e., $\lambda_i$ satisfying above restriction) are used in our protocol, let $\rho^{ij}_f$ denotes the state shared between Alice and Charlie depending on any specific output $b_{ij}$(as already discussed before). There exist some absolutely Bell-CHSH local states belonging to the family of Gisin states(Eq.(\ref{gis1})) for which, irrespective of Bob's output, non-absolutely Bell-CHSH local states are generated in the protocol if Bob obtains output $|\phi^+\rangle (b_{00})$ or $|\phi^-\rangle(b_{01})$. For instance, when $\lambda_1=\lambda_2$, generation of non absolute Bell-CHSH local states is shown in Fig.(\ref{abs1}).\\
We then considered generating identical copies of the state $\rho_1$(say). Interestingly non absolute locality was generated(for Bob obtaining $b_{00}$ or $b_{01}$) in the measurement phase of the protocol in this case also(see Fig.(\ref{abs2})).
\begin{center}
\begin{figure}[htb]
	\centering
	\includegraphics[width=2.5in]{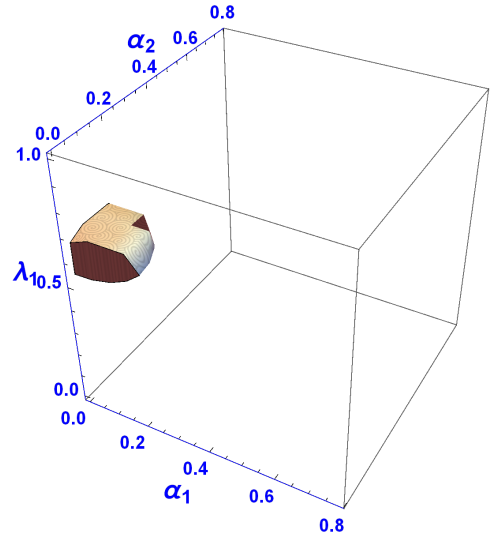}\label{abs1}
\caption{\emph{Here $\lambda_2=\lambda_1.$ Shaded region gives a restricted set of state parameters $\lambda_1,$ $\alpha_1$ and $\alpha_2$ of the states $\rho_f^{00}$ or $\rho_f^{01}$ for which starting from two non identical copies of absolutely local Gisin states, non absolutely Bell-CHSH local states are generated in the measurement phase of the protocol.}}
\label{abs1}
\end{figure}
\end{center}
Till now we discussed our findings related to generation from absolutely Bell-CHSH local states. In the context of exploring nonlocal feature of any quantum state it is important to distinguish states which have a local hidden variable(LHV) model, i.e, are local in terms of hidden variable theory as there exist entangled states which lack a LHV model but fail to show violation of Bell-CHSH inequality \cite{CHSH69}.\\
For instance, consider the family of states:

\begin{eqnarray}\label{l1}
 \varrho^L=\gamma(\cos^2\beta|01\rangle\langle01|+\sin^2\beta|10\rangle\langle10|\nonumber\\-\sin\beta\cos\beta(|10\rangle\langle01|+|01\rangle\langle10|))\nonumber\\+(1-\gamma)|00\rangle\langle00|\nonumber\\\beta\in[0,\frac{\pi}{2}],\,\gamma\in[0,1].    \end{eqnarray}

This class of states is entangled for any $\beta\neq 0,\frac{\pi}{2}.$ However this family has an LHV model(upto projective measurements) for any $\gamma\leq\frac{1}{1+\sin(2\beta)}$\cite{localmodel}. For $\gamma=\frac{1}{2},$ the corresponding states from this family are absolutely local. So for any possible value of $\beta$, the states are both absolutely Bell-CHSH local and also have a LHV model. We now explore whether nonlocal behavior of such seemingly useless states can be exploited via our protocol.
\subsection{Nonlocality from absolutely Bell-CHSH local states having LHV model}
In the standard Bell scenario, states admitting a LHV occupy a significant position. Any statistics arising out of them can be efficiently simulated by shared randomness. In what follows below, we show that our prescription can generate nonlocality from absolutely Bell-CHSH local states admitting a LHV.\\
\indent Let each of $S_1$ and $S_2$ generates an absolutely Bell-CHSH local state  having LHV model from the family of of states $\varrho_L$(Eq.(\ref{l1})). To be specific let $S_i(i=1,2)$ generate:

\begin{eqnarray}\label{l2}
   \varrho^L_i=\frac{1}{2}(\cos^2\beta_i|01\rangle\langle01|+\sin^2\beta_i|10\rangle\langle10|\nonumber\\-\sin\beta_i\cos\beta_i(|10\rangle\langle01|+|01\rangle\langle10|)+|00\rangle\langle00|)\nonumber\\\beta_i\in[0,\frac{\pi}{2}].
\end{eqnarray}

\begin{center}
\begin{figure}[htb]
\begin{tabular}{|c|c|c|}
\hline
\subfloat[$\rho_f^{00}$ or $\rho_f^{01}$ ]{\includegraphics[trim = 0mm 0mm 0mm 0mm,clip,scale=0.2]{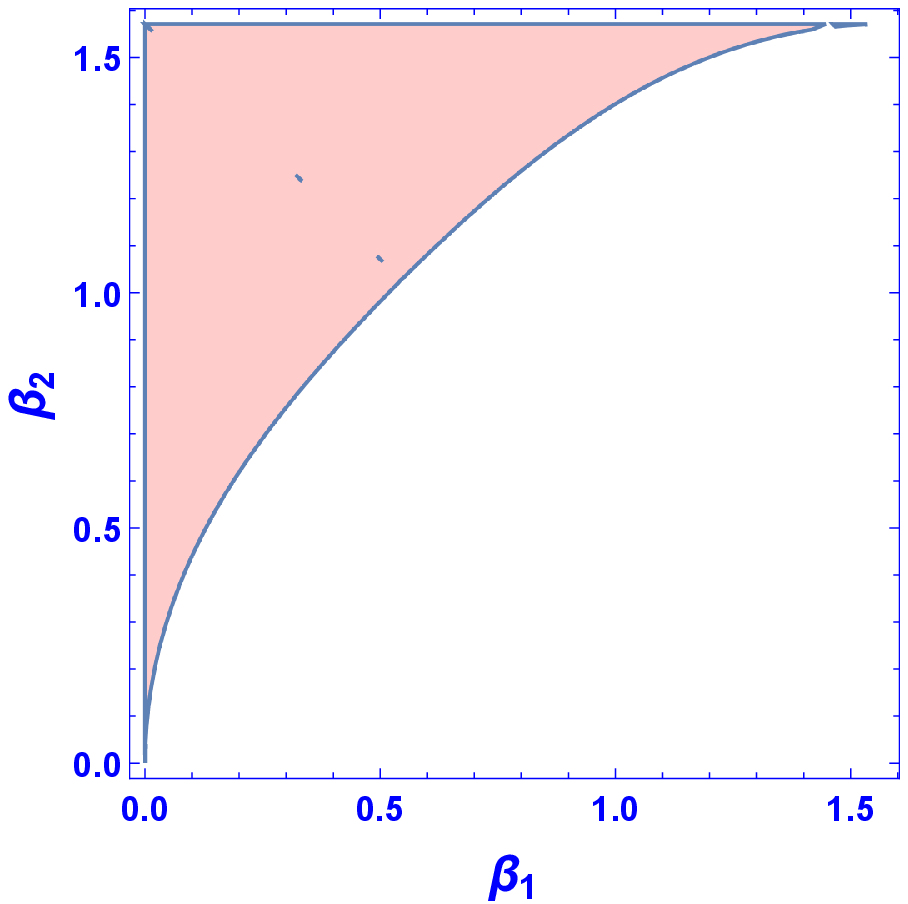}} &
\subfloat[$\rho_f^{10}$ or $\rho_f^{11}$]{\includegraphics[trim = 0mm 0mm 0mm 0mm,clip,scale=0.2]{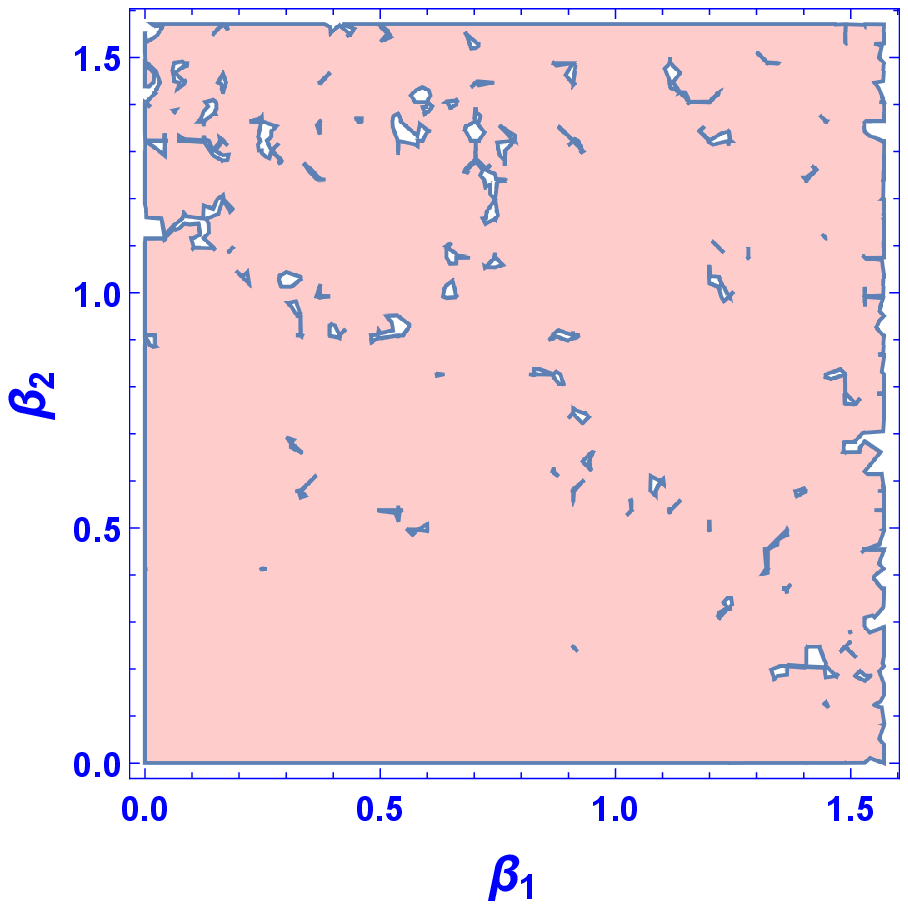}}&
\subfloat[$\rho_f^{ij}$ $i,j\in\{0,1\}$]{\includegraphics[trim = 0mm 0mm 0mm 0mm,clip,scale=0.2]{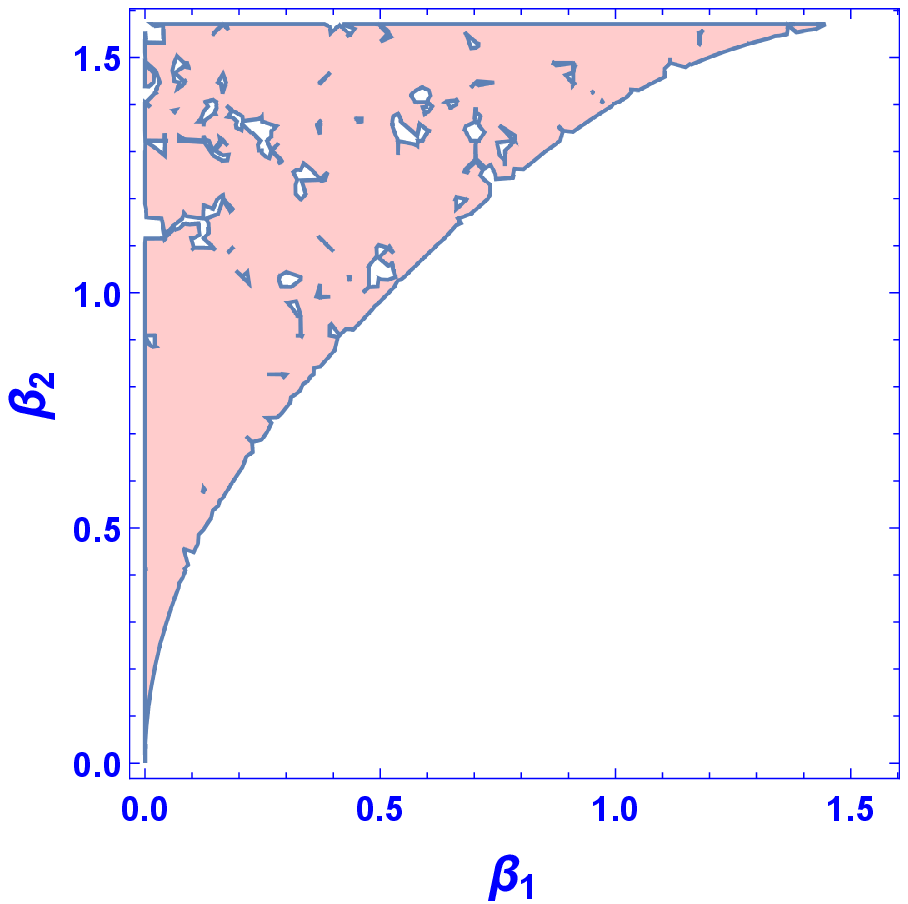}}\\
\hline
\end{tabular}
\caption{\emph{Subfigures(a) and (b) give restrictions over state parameters $\beta_1$ and $\beta_2$ of the two absolutely local copies of the state(Eq.(\ref{l1})) for which non absolute local states are generated in the measurement phase of the protocol. While revelation is observed for Bob obtaining $|\phi^{\pm}\rangle$ in subfigure(a), same is obtained for Bob obtaining output $|\psi^{\pm}\rangle$ in subfigure(b). Non empty intersection of the regions in subfigure(a) and subfigure(b) is given in subfigure(c) which in turn indicates deterministic generation of non absolute locality, i.e., generation independent of Bob's output.}}\label{abs5}
\end{figure}
\end{center}

\begin{figure}[htb]
	\centering
	\includegraphics[width=2.5in]{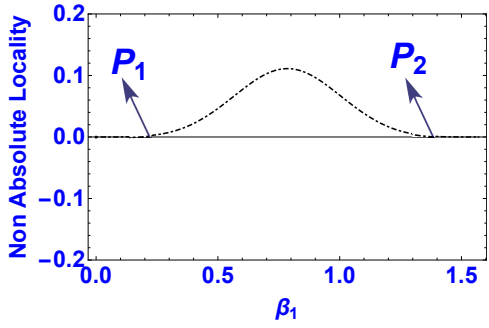}
\label{2}
	\caption{\emph{Portion of the curve above $\beta_1$ axis represents revelation of non absolute local states at the end of our protocol(measurement phase). Intercepted length $P_1P_2$ of the $\beta_1$ axis gives the restriction over the state parameter $\beta_1$ for which generation of nonlocality with some non zero probability is obtained starting from absolutely Bell-CHSH local states having LHV model(upto projective measurements).}}
	\label{abs6}
\end{figure}

Non absolutely local states are generated at the end of the protocol for any output of Bob(see Fig.(\ref{abs5})).
Again if both the sources generate identical copies of the state $\varrho_1^L$(say), then non-absolutely local states are generated probabilistically if Bob obtains output $|\psi^{\pm}\rangle.$(see Fig.(\ref{abs6})). Probability of success in each case($|\psi^{+}\rangle$ or $|\psi^{-}\rangle$) is $\frac{7+\cos(4\beta_1)}{32}.$ Clearly by using identical copies of any absolutely Bell-CHSH local state(having LHV model) of this family(Eq.(\ref{l1})) in our protocol, non-absolutely Bell-CHSH local states are produced with some non zero probability.
One should note that in each of the above demonstrations, the swapping protocol converts a state into a state which is not absolutely Bell-CHSH local. A suitable global unitary needs to be applied on the resultant state to generate nonlocality as already mentioned in the description of our protocol in sec. \ref{cs1}.
\vskip-2cm
\section{Conclusions} \label{concl}
\indent Bell nonlocality occupies a paramount position in the foundational aspects of quantum theory. It is a significant ingredient in various quantum information processing tasks. However, in practical scenarios such resource can be destroyed due to environmental influence.\\
\indent Nonlocality can be destroyed through environmental interactions. A pertinent question is then related to finding measures to retrieve nonlocality. This question motivates our present work. In this work, we have come across a situation where the drastic effect of environment(in the form of a amplitude damping channel) can lead a nonlocal state to be absolutely Bell-CHSH local. Absolutely Bell-CHSH local states are precisely those from which nonlocality cannot be generated even by using global unitary operations. Our work here proposes a prescription through which one can retrieve nonlocality. The prescription relies on the entanglement swapping protocol. Using the proposal one can turn a state which is absolutely Bell-CHSH local to a state which is no longer absolutely Bell-CHSH local. One can then use a global unitary operation on the resultant state to generate nonlocality. The X class of states, which contains within it several important classes of states like the Werner and Gisin states, have been taken as the input in our protocol. We have shown successfully that nonlocality can be generated from otherwise seemingly useless states(in terms of nonlocality).\\
\indent The present work leads to useful directions of future research. We have investigated the effect of nonlocality breaking channels and the subsequent retrieval of nonlocality in the purview of the Bell-CHSH inequality in two qubits. Extensions of the work to other Bell inequalities in two qubits can be interesting. The scenario in multiqubit systems can also turn to be a significant enquiry.
\vskip-2cm
\section*{Acknowledgement}
We thank Prof. Guruprasad Kar for his enriching inputs. AM acknowledges support from the CSIR project 09/093(0148)/2012-EMR-I. DS acknowledges finacial support from DST-SERB and DSA-SAP.

\end{document}